\documentclass[]{spie}  

\usepackage{amsmath,amsfonts,amssymb,xspace}
\usepackage{graphicx}
\usepackage[colorlinks=true, allcolors=blue]{hyperref}

\newcommand{\mum}{\mbox{{\usefont{U}{eur}{m}{n}{\char22}}m}\xspace}

\newcommand{\SubItem}[1]{
    {\setlength\itemindent{15pt} \item[-] #1}
}

\title{AO3k + SCExAO: on-sky wavefront quality and demonstration of novel WFS techniques with the double XAO system}

\author[a,b]{Julien Lozi}
\author[c,a]{Kyohoon Ahn}
\author[d,a]{Vincent Deo}
\author[a,b,e,f]{Olivier Guyon}
\author[e,a]{Sandrine Juillard}
\author[a]{Yoshito Ono}
\author[a,b]{Garima Singh}
\author[g,h,a]{S\'{e}bastien Vievard}
\affil[a]{Subaru Telescope, National Astronomical Observatory of Japan, National Institutes of Natural Sciences (NINS), 650 North A`oh\={o}k\={u} Place, Hilo, HI 96720, United States}
\affil[b]{Astrobiology Center of NINS, 2 Chome-21-1, Osawa, Mitaka, Tokyo, 181-8588, Japan}
\affil[c]{Korea Astronomy and Space Science Institute Daedeokdae-ro 776, Yuseong-gu Daejeon 34055, Republic of Korea}
\affil[d]{Optical Sharpeners, 04119 Manosque, France}
\affil[e]{Steward Observatory, University of Arizona, Tucson, AZ 87521, United States}
\affil[f]{College of Optical Sciences, University of Arizona, Tucson, AZ 87521, United States}
\affil[g]{Space Science and Engineering Initiative, College of Engineering, University of Hawai‘i, Hilo, HI 96720, USA}
\affil[h]{Institute for Astronomy, University of Hawaii, Hilo, HI 96720, USA}

\authorinfo{Further author information: (Send correspondence to J.L.)\\J.L.: E-mail: lozi@naoj.org, Telephone: 1 808 934 5949}

\begin{document} 
\maketitle

\begin{abstract}
The Subaru Coronagraphic Extreme Adaptive Optics (SCExAO) system, fed by its upstream 3000-actuator “woofer” (AO3k), serves both as a platform for high contrast imaging (HCI) technology maturation and as a science instrument for imaging, spectroscopy, and polarimetry of exoplanets and disks. SCExAO operates in the visible and near-IR and offers a wide choice of instrument configurations.
Over the last year, AO3k/SCExAO underwent significant upgrades to bring improved capabilities and support new developments, all while easing science operations. The new configuration features a beam switcher so that light can be shared between several instrument modules. The system is evolving toward a tighter integration between multiple WFSs and AO stages of correction, with the first stage (AO3k) providing visible and nearIR WFSing, as well as laser tomography.

AO3k+SCExAO has been fully operational since October 2025, demonstrating very high stability on-sky, even in bad seeing conditions up to 2”. Having two XAO in series allows us to deploy advanced wavefront control techniques optimized for high-contrast imaging (e.g. speckle nulling, EFC, Coronagraphic LOWFS, Fast and Furious) on the second-stage XAO loop, as AO3k by itself delivers high-contrast PSFs already. Areas of active ongoing research include use of photonic devices for spectrally dispersed interferometric sensing, PSF reconstruction from WFS telemetry, and non-linear sensors (focal plane and curvature). Recent upgrades to the computer infrastructure are aimed at supporting these R\&D efforts and providing a rich collaborative environment for experimentation. 

In this paper, we will present on-sky high-contrast performance characterization of AO3k, AO3k+SCExAO, and on-sky demonstrations of novel wavefront control techniques to improve the contrast behind the coronagraph.

\end{abstract}


\section{INTRODUCTION}
\label{sec:intro}  

The last couple of years were exciting for the AO instrumentation at Subaru. We improved the facility adaptive optics AO188\cite{Takami2006, Hayano2006} to AO3k, by upgrading the deformable mirror (DM) to ALPAO's 3224-actuator DM\cite{Lozi2024}. We also added a Near-Infrared Wavefront Sensor (NIRWFS)\cite{Lozi2022,Lozi2024} and a non-linear Curvature Wavefront Sensor (nlCWFS)\cite{Ahn2023}. Finally, we also added a polychromatic source, simulating a star going through the pupil of Subaru for alignments and WFS calibration. 

The facility adaptive optics feeds several instruments downstream:

\begin{itemize}
    \item The facility instrument IRCS (Infrared Camera and Spectrograph), providing imaging (y- to M'-bands), polarimetry and spectroscopy (zJ- to L-band) capabilities.
    \item The PI extreme-AO platform SCExAO (Subaru Coronagraphic Extreme Adaptive Optics)\cite{Jovanovic2015}, feeding several science modules in visible (R- and i-band) and NIR (y- to K-band), including the integral field spectrograph CHARIS, the MKID Exoplanet Camera (MEC), the visible differential polarimetric imager VAMPIRES\cite{Lucas2024b} and the newly open for science FIRST-PL\cite{Vievard2024b}.
    \item the PI instrument IRD (Infrared Doppler spectrograph), a NIR (y- to H-band) fiber-fed high resolution spectrograph. IRD can be fed directly with multi-mode fibers from AO188, or with single-mode fibers from SCExAO (REACH module)\cite{Kotani2018}.
\end{itemize}

In addition to these instruments, the newly installed Nasmyth Beam Switcher (NBS) allows to add a visitor port for more AO instruments in the future. 

Upgrading AO188 to AO3k means that now, combined with SCExAO, we have two Extreme-AO back-to-back. This should allow us to let AO3k focus on correcting the atmospheric turbulence, while SCExAO can use its DM to focus on more advanced and more precise wavefront correction. For example, we want to use speckle control to create and maintain a high contrast region in the focal plane, to improve the detection of faint exoplanets. Finally, we want to test technologies necessary to prepare for the next generation of giant segmented telescopes, both ground-based and in space, like the Thirty Meter Telescope (TMT) and the Habitable Worlds Observatory (HWO).

In this paper, we will present the final configuration of the AO instruments at Subaru in Sec.~\ref{sec:config}, especially the upgrades done on the NIRWFS in Sec.~\ref{sec:nirwfs_upgrades}. Then we will present in Sec.~\ref{sec:lwe} the current limitations in the performances, mostly due to the low-wind effect (LWE). Finally, we will discuss the past and current efforts to perform speckle control on-sky in Sec.~\ref{sec:speckle_control}.

\section{FINAL CONFIGURATION OF THE AO INSTRUMENTS OF SUBARU}
\label{sec:config}

\subsection{Phase II of the NasIR \& Ao3k Upgrades: Nasmyth Beam Switcher}
\label{sec:phaseII}

 Several upgrades were planned over the last couple years for the AO instrumentation of Subaru. The facility AO, AO188, was upgraded to AO3k, with a new 3000-actuator deformable mirror from ALPAO, and new wavefront sensors matching the actuator count. An optical relay, called the Nasmyth Beam Switcher (NBS)\cite{Hattori2024} was also added in summer of 2025, allowing us to have several instruments behind the facility AO, without requiring us to crane each instrument separately. The instruments rest on a common platform, a steel frame that improves the stability of the instrumentation behind AO3k.

\begin{figure} [t]
\begin{center}
\begin{tabular}{c}
\includegraphics[width=0.8\textwidth]{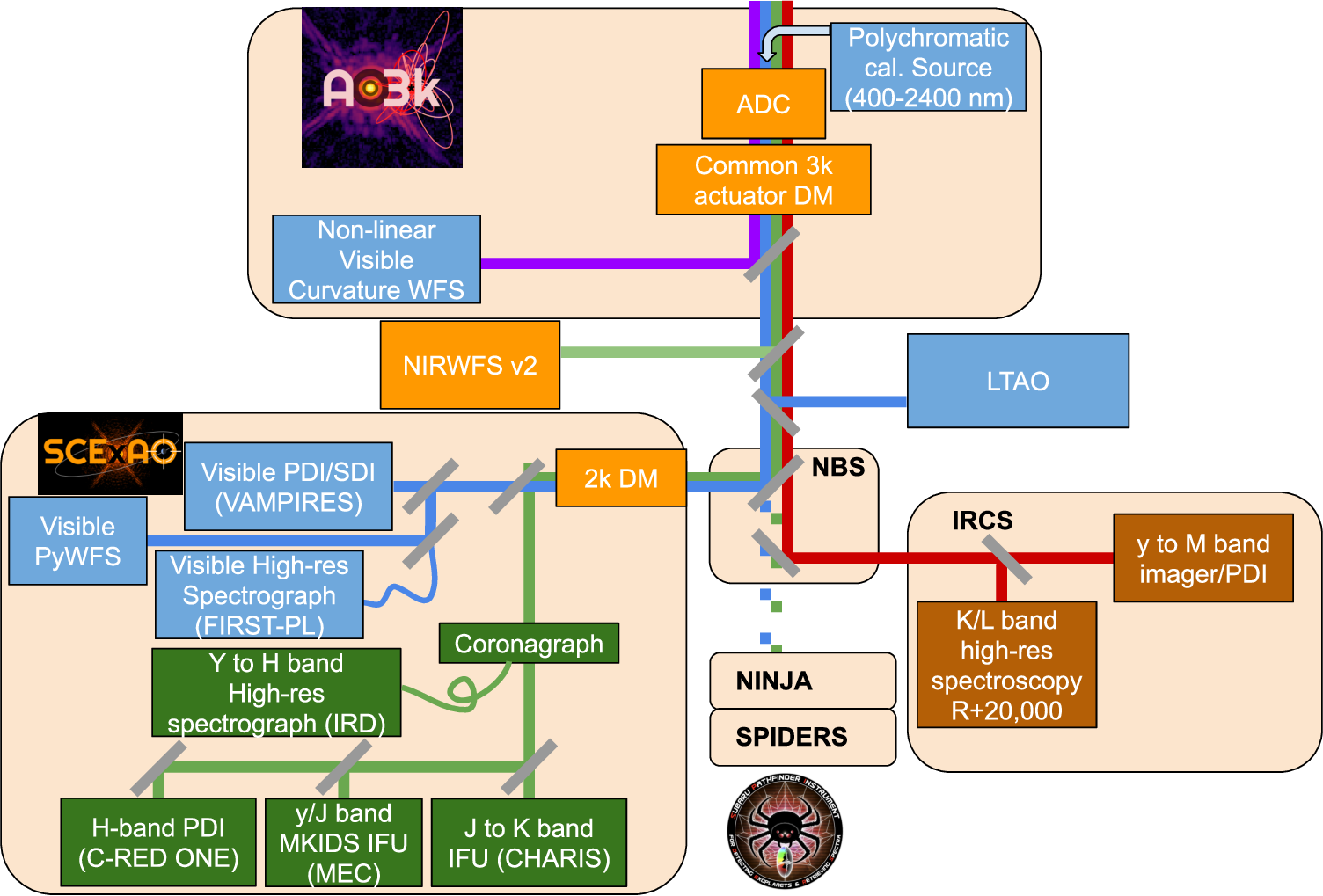}
\end{tabular}
\end{center}
\caption{Final configuration of the IR Nasmyth platform of Subaru: The light from the telescope goes through a first layer of extreme correction with AO3k, including an upgraded NIRWFS and LTAO system, then the light goes through the NBS, which relays the beam to several ports. On one side is Suabru's facility instrument IRCS, and on the other side is SCExAO. A couple of visitor ports are left for new instruments, with SPIDERS using one of them as a tech demo for less than a year, and the spectrograph NINJA arriving this summer to replace it.}
\label{fig:ao3k_phaseII} 
\end{figure} 

Figure~\ref{fig:ao3k_phaseII} presents the final configuration after the installation of theNBS. It was designed to switch rapidly between up to 4~instruments without requiring craning as it was done behind AO188/AO3k.The light can potentially be split between 2~instruments (e.g. SCExAO and IRCS, or SCExAO and SPIDERS) with a dichroic beamsplitter.
    
In this final phase of the NasIR upgrades, the gap between AO3k and the NBS is filled with a platform supporting the laser tomography AO (LTAO) currently in development\cite{Akiyama2020} called ULTIMATE-START, a precursor of the future MCAO ULTIMATE-SUBARU. This platform also allowed us to move the NIRWFS of AO3k to a larger area, providing more space for a few upgrades.

Two new instruments were added thanks to the NBS: SPIDERS\cite{Thompson2024}, a high-contrast high-resolution spectro-imaging test platform, visiting from Fall of 2025 to Summer of 2026, and NINJA\cite{Tokoku2022}, a multi-purpose visible-NIR slit spectrograph that will be installed in August 2026 in place of SPIDERS.

\begin{figure} [b]
\begin{center}
\begin{tabular}{c}
\includegraphics[width=0.75\textwidth]{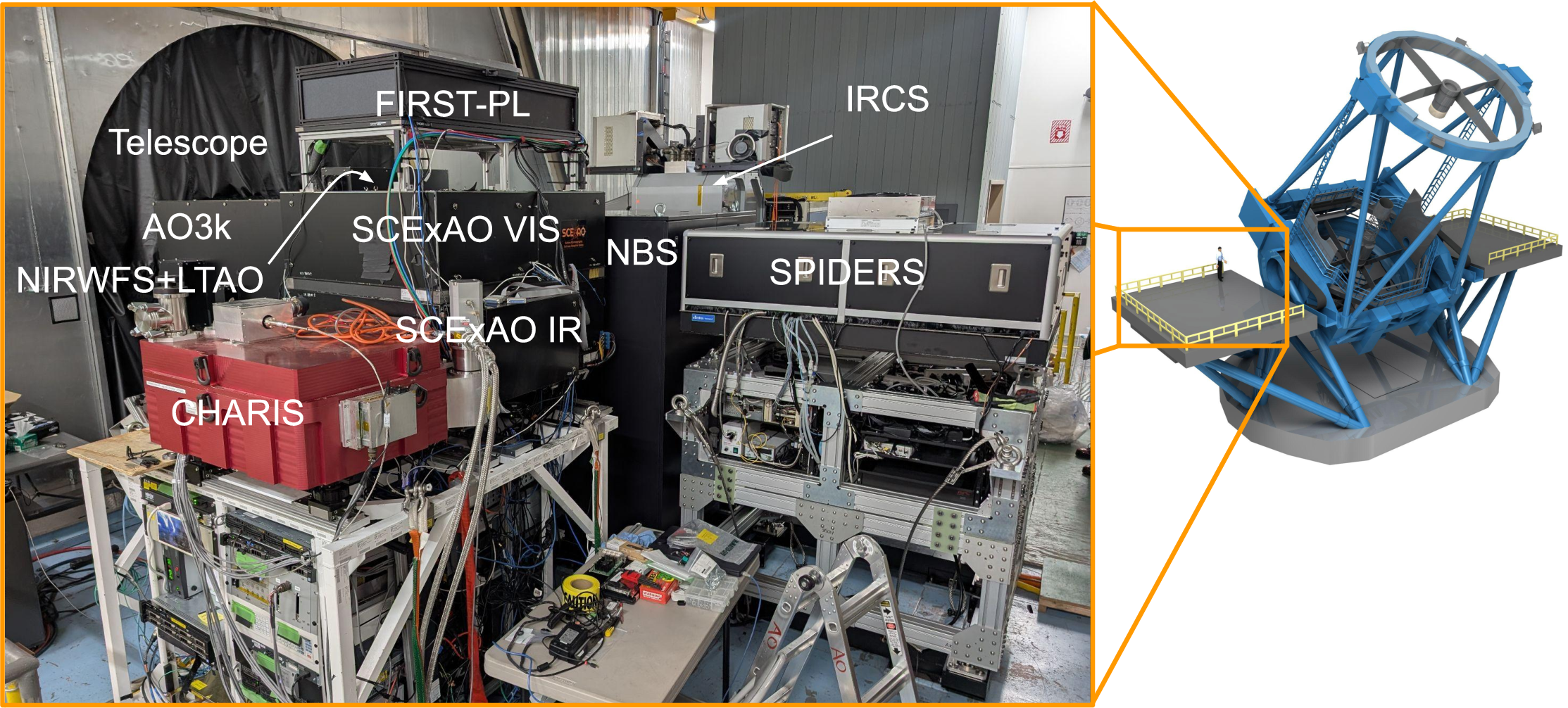}
\end{tabular}
\end{center}
\caption{Picture of the instrumentation on the NasIR of Subaru, taken when SPIDERS was visiting. It was removed just before SPIE, to leave the space to the spectrograph NINJA.}
\label{fig:scexao_subaru} 
\end{figure} 

Figure~\ref{fig:scexao_subaru} shows the instrument configuration on NasIR while SPIDERS was visiting. The NIRWFS \& LTAO platform, the NBS, IRCS and SPIDERS/NINJA are resting on the common platform, because of the beam height change introduced by the NBS, while SCExAO is still resting directly on the NasIR platform. The new location of SCExAO actually introduced another challenge: we noticed stronger vibrations in the science frames for some telescope orientations. We confirmed that these vibrations were not seen by AO3k and the NIRWFS, and were unique to SCExAO's location. These vibrations are telescope vibrations\cite{Lozi2018b} that are transferred to SCExAO's support frame via the NasIR platform plates. Although they are mostly managed by SCExAO's AO loop, we are looking into reinforcing the supports below SCExAO's feet to reduce the coupling of these vibrations.

\subsection{AO3k/NIRWFS upgrades}
\label{sec:nirwfs_upgrades}

\begin{figure} [b]
\begin{center}
\begin{tabular}{c}
\includegraphics[width=0.8\textwidth]{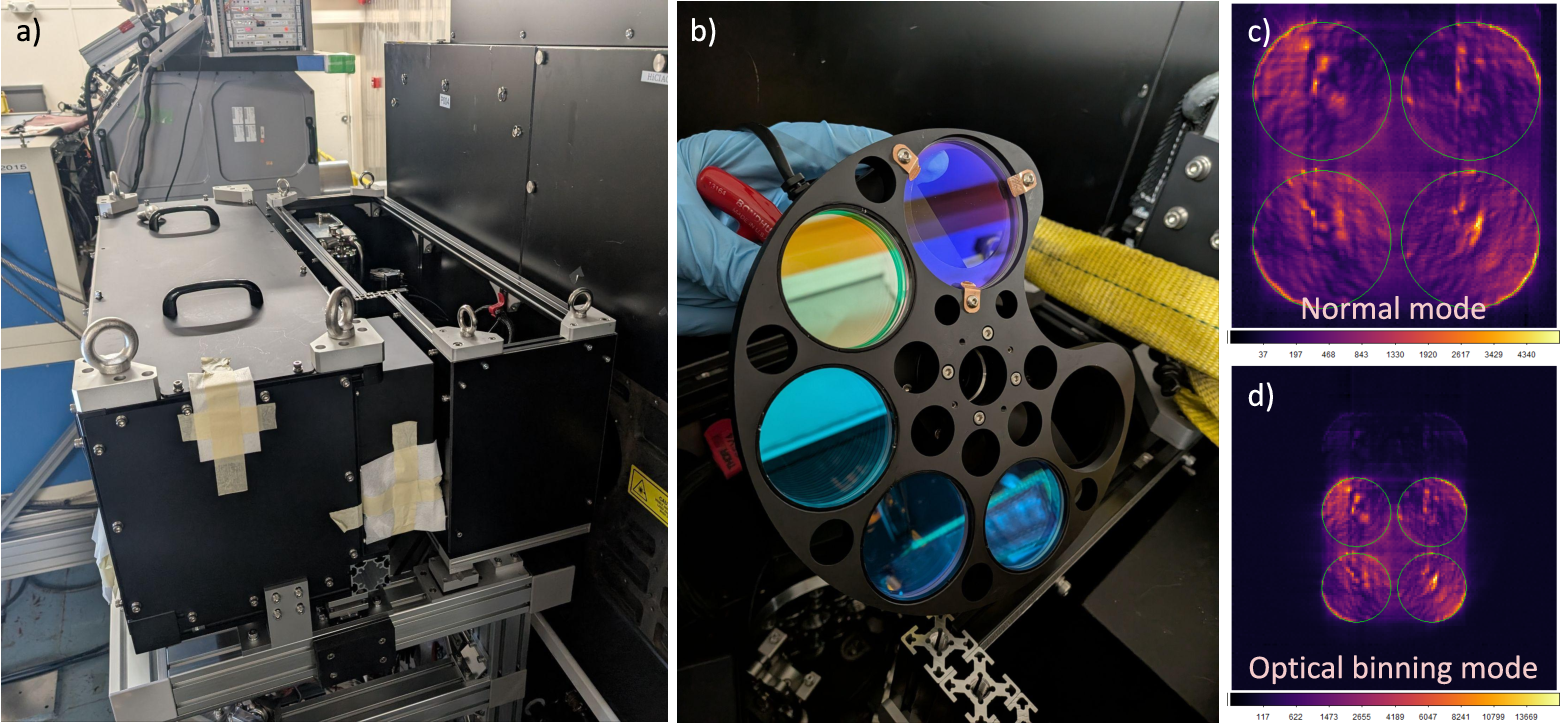}
\end{tabular}
\end{center}
\caption{(a) NIRWFS v2 (open box on the right) installed on the LTAO platform, next to the truth SHWFS (closed box on the left). (b) New NIRWFS pickoff beamsplitter wheel. (c) Original PyWFS mode of the NIRWFS (high-res mode). (d) Optical binning mode, where each pupil is 2~times smaller (low-res mode). }
\label{fig:nirwfs_v2} 
\end{figure} 

With the addition of the NBS, the NIRWFS was moved to the output port of AO188/AO3k, on the LTAO platform. Figure~\ref{fig:nirwfs_v2} (a) shows the upgraded NIRWFS, next to the truth Shack-Hartmann WFS of ULTIMATE-START. This new location allowed us to improve on the design of the WFS, thanks to a larger available footprint. For NIRWFS v2, we added a few useful features:

\begin{itemize}
\item \textbf{A dichroic beamsplitter wheel:} this motorized wheel (see fig~\ref{fig:nirwfs_v2} (b)) allows us to change dichroics when needed, even between targets in the same observation. This is an improvement from the previous design, where only one dichroic was available per observing night. The existing pickoffs are as follows:
	\SubItem{\textbf{K-band dichroic}, sending yJH to the NIRWFS, and K-band to the science path. This is mostly for IRCS, but can be used for CHARIS K-band.}
	\SubItem{\textbf{yJH50 dichroic}, sending 50\% of the light in yJH to the NIRWFS, and the other 50\%, as well as all of K-band and visible light between 600 and 900~nm to the science path. This is the main dichroic used by SCExAO, to feed both VAMPIRES and CHARIS.}
	\SubItem{\textbf{yJH90 dichroic}, sending 10\% of the light in yJH to the NIRWFS, and the other 90\%, as well as all of K-band and visible light between 600 and 900~nm to the science path. This dichroic is ideal for bright targets, sending most of the NIR light to the science path.}
	\SubItem{\textbf{visHK dichroic}, sending all of y and J-band to the NIRWFS, and the all of H-band, K-band and visible light between 600 and 900~nm to the science path.}
	\SubItem{\textbf{L-band dichroic}, sending yJHK to the NIRWFS and L-band light, as well as visible light to the science path. This is a new addition, mostly aimed for L-band imaging with IRCS. The NIRWFS is not sensitive to K-band at this point, but a modification of the C-RED ONE is envisioned to allow K-band wavefront sensing in the future.}
\item \textbf{A NIR acquisition camera:} A small NIR camera is added inside the instrument, to acquire the targets more efficiently. This camera reduces the overhead between targets, and allows to measure the seeing routinely between targets. Its software integration is on-going.
\item \textbf{An adaptive field stop:} The NIRWFS struggles to close the loop if one or several other bright stars are present in the field of view of the sensor. This field stop blocks the light from these stars to reach the camera. Its size can be adjusted depending on the position of the other stars.
\item \textbf{An optical binning mode:} Another pupil lens configuration is added to reduce the size of the four pupils on the camera by a factor two:  34~pixels across (see fig~\ref{fig:nirwfs_v2} (d)) instead of 68~pixels across for the original mode (see fig~\ref{fig:nirwfs_v2} (c)). This means that the light is more concentrated by a factor 4 on each pixel, which increases the magnitude limit by almost two magnitudes. However, in this configuration, also called low-resolution or LR configuration, we cannot control as many modes on the DM, so the performance will not be as high as the original PyWFS configuration, also called high-resolution or HR configuration.
\end{itemize}

\begin{figure} [b]
\begin{center}
\begin{tabular}{c}
\includegraphics[width=0.8\textwidth]{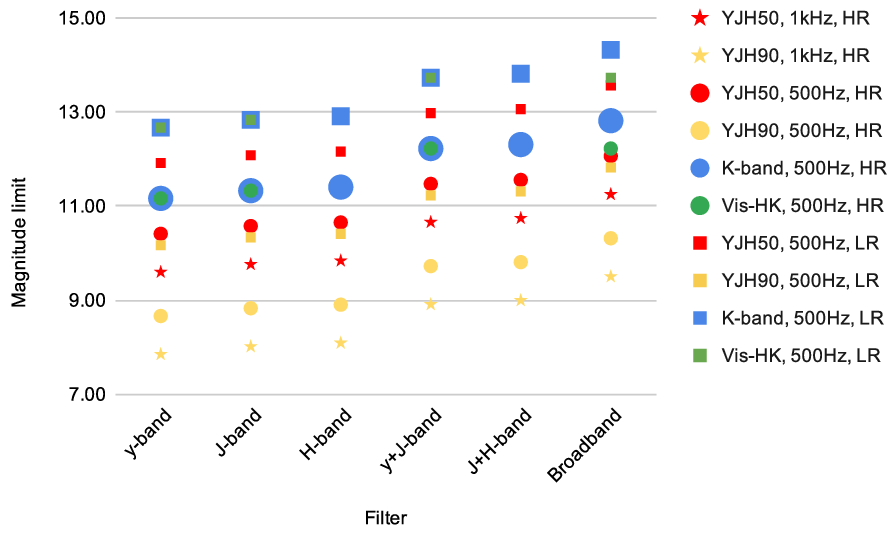}
\end{tabular}
\end{center}
\caption{Guide star magnitude limit. The frequencies listed in the legend are the AO loop frequency, and the mode (HR for high-res, or LR for low-res). Correction at 500 Hz might be worse than at 1 kHz depending on the atmospheric conditions (seeing, coherence time). These values assume guide stars with a flat spectrum. For red stars, the H-band limit is more accurate than the Broadband one.}
\label{fig:nirwfs_maglim} 
\end{figure} 

Figure~\ref{fig:nirwfs_maglim} shows the magnitude limits of the NIRWFS v2, for the various dichroic pickoffs, bands and modes. These values assume a flat spectrum for the guide star, so typically A-type stars. For redder objects, it is better to consider the H-band magnitude limit instead of the broadband one, since most of the contribution to the WFS will be in H-band. In broadband mode, with the K-band dichroic at 500~Hz, we can reach stars with H-magnitudes close to 13, while the new low-resolution mode gets us close to 14.5. 

\section{Low-Wind Effect or Island Effect: a Strong Limitation}
\label{sec:lwe}

We presented the on-sky performances of AO3k and SCExAO with the first version of the NIRWFS in \cite{Lozi2024}. The upgrade of the NIRWFS did not impact the performance in the HR mode, and the analysis of the data in LR mode is still on-going. The NBS only added about 30~nm of low-order aberrations to the various ports, which is easily managed by SCExAO's DM. Unfortunately, some degradation in vibrations was observed as described previously. This can put some unnecessary strain on SCExAO's PyWFS, so that is why we are looking into reinforcing the mounting points of the instrument on the platform.

\begin{figure} [b]
\begin{center}
\begin{tabular}{c}
\includegraphics[width=0.6\textwidth]{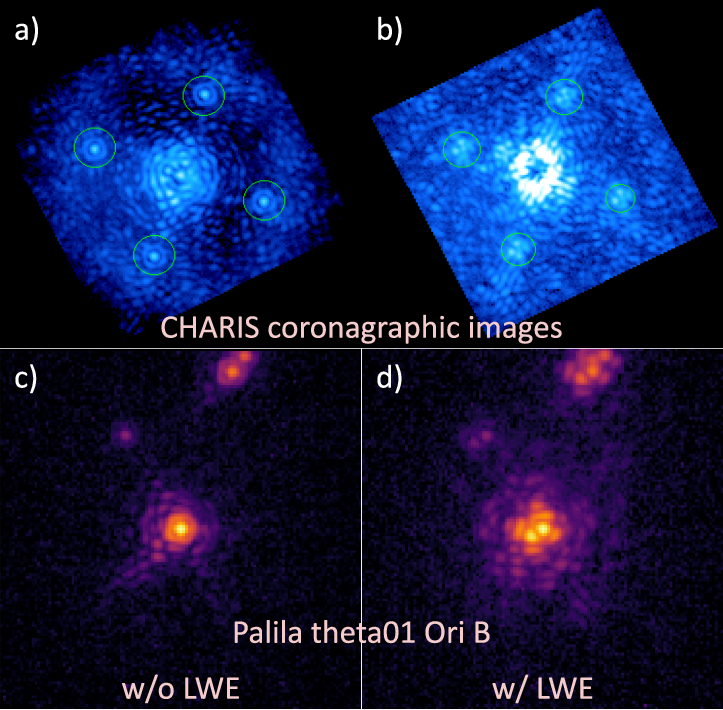}
\end{tabular}
\end{center}
\caption{Impact on the low-wind effect on science images. (a) and (b) CHARIS coronagraphic image, with astrometric grid, without and with LWE respectively. The speckle field is heavily impacted by LWE, and leakage around the coronagraphic mask gets worse as well. (c) and (d) Palila C-RED 2 camera images on the multiple star system theta01 Ori B, without and with LWE respectively. Each star splits in two in that case, which makes astrometry difficult.}
\label{fig:nlwe_images} 
\end{figure} 

The other main issue still encountered by AO3k and SCExAO is the infamous low-wind effect (LWE), or island effect, which affects us 30 to 50\% of the time. LWE comes from the temperature difference between the telescope spiders and the ambient air. Due to radiative cooling, the spiders can be colder than the surrounding air, which in a low-wind environment, gets cooled down itself. This creates a variation of air temperature around the spiders, that translates into a phase step between the two sides of the spiders. Modulated PyWFS are not the best at sensing phase differences between the parts of the pupils, also called petal modes, which means that they are not well corrected in case of LWE. Even when LWE is not present, a PyWFS can introduce some petal modes randomly. These modes end up splitting the PSF into 2 to 4 lobes, as seen in Fig~\ref{fig:nlwe_images}, degrading significantly the contrast in coronagraphic images, and making astrometric analysis difficult.

We are working on several solutions to correct the petal modes: Fast \& Furious showed promising results on non-coronagraphic images\cite{Bos2020}, 2 Fast 2 Furious\cite{Bottom2023}, which would work on coronagraphic images, Lyot-stop Low-Order Wavefront Sensor using machine learning, which was demonstrated on-sky just before this conference (Garima Singh et al., these proceedings), Zernike Wavefront Sensor\cite{Ndiaye2018}, Non-redundant masking\cite{Deo2024b}, wavefront sensing with photonic devices\cite{Rossini2024}, etc. Most of these were demonstrated in laboratory conditions, introducing petal modes artificially, and some of them were demonstrated on-sky in real conditions. Unfortunately, there is no easy and unique solution for every observing mode.

\section{SPECKLE CONTROL}
\label{sec:speckle_control}

\subsection{History of Speckle Control with SCExAO}
\label{sec:history}

\begin{figure} [b]
\begin{center}
\begin{tabular}{c}
\includegraphics[width=0.97\textwidth]{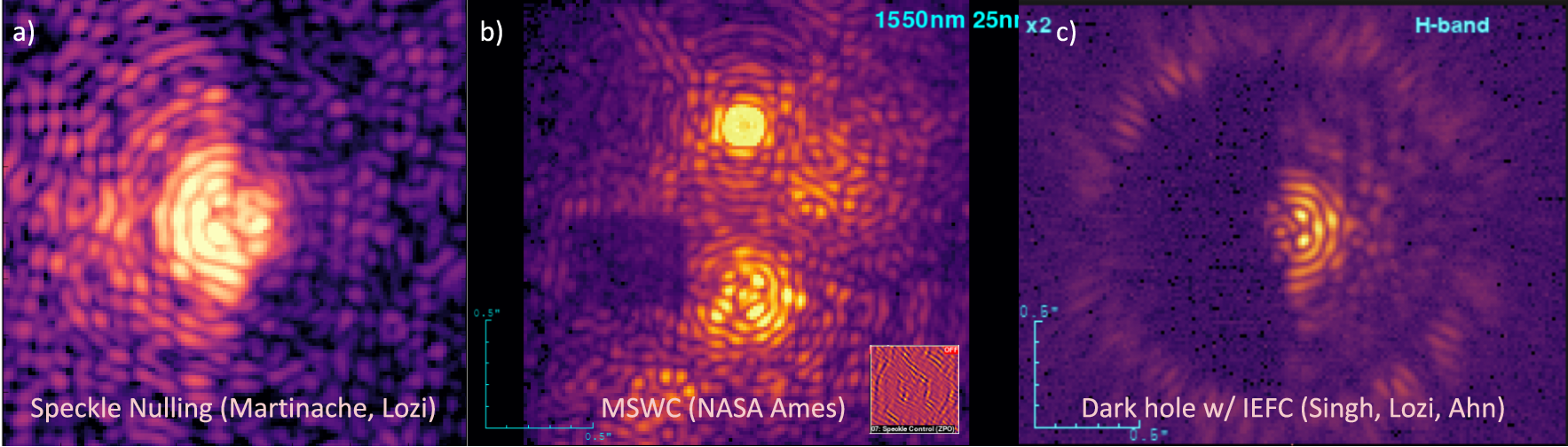}
\end{tabular}
\end{center}
\caption{Various speckle control algorithms tested on SCExAO over the years: (a) Classical speckle nulling was used first to dig dark hole regions on the internal source. Static maps were applied on-sky, providing small improvements on contrast. (b) Multi-Star Wavefront Control was demonstrated on our internal source, by sequentially digging the same region for both stars, and summing up the contributions in post-processing. (c) IEFC was implemented on SCExAO to create dark holes and demonstrate their stabilization with Linear Dark Field Control. Now IEFC will be used on-sky.}
\label{fig:speckle_control} 
\end{figure} 

SCExAO, as a high-contrast imaging testbed, tested several speckle control algorithms over the years. We started with speckle nulling to create D-shape dark holes in coronagraphic and non-coronagraphic images in NIR\cite{Martinache2014} (see fig~\ref{fig:speckle_control} (a)). We created dark hole maps for the DM with the internal source, and tested applying these maps on-sky, with some improvements in the contrast. Running speckle nulling on-sky was never successful, due to high wavefront residuals from AO188, that SCExAO's DM had to correct. 

During COVID lockdowns, the team at NASA Ames tested their Multi-star Wavefront Control (MSWC) algorithm remotely on SCExAO\cite{Pluzhnik2021}, using our internal source (see fig~\ref{fig:speckle_control} (b)). They dug a dark hole for the main star behind the coronagraph, then using the same DM, dug the dark hole for the off-axis companion simulated by physically moving the source in the field. The final result was obtained by combining both images in post-processing. This algorithm was not tested on-sky.

More recently, we used Implicit Electric Field Conjugation (IEFC)\cite{Haffert2023} to dig dark holes in laboratory settings, and combined it with Linear Dark Field Control (LDFC) to demonstrate the stabilization of the high-contrast region using the bright light on the other side\cite{Ahn2023b}. 

So far, on-sky testing was unsuccessful, because the SCExAO DM was doing a lot to correct residual atmospheric disturbance, and the only way to send DM correction in this case is to modify the reference point of the visible PyWFS. In theory, now that AO3k is working, the first layer of wavefront control already provides extreme-AO correction, which should allow the second layer of wavefront control to focus on speckle correction. SPIDERS demonstrated this fact actually (Christian Marois, these proceedings), by digging dark holes on-sky with their self-coherent camera on top of the AO3k correction. 

\subsection{IEFC Used to Flatten the DM}
\label{sec:superflat}

\begin{figure} [t]
\begin{center}
\begin{tabular}{c}
\includegraphics[width=0.7\textwidth]{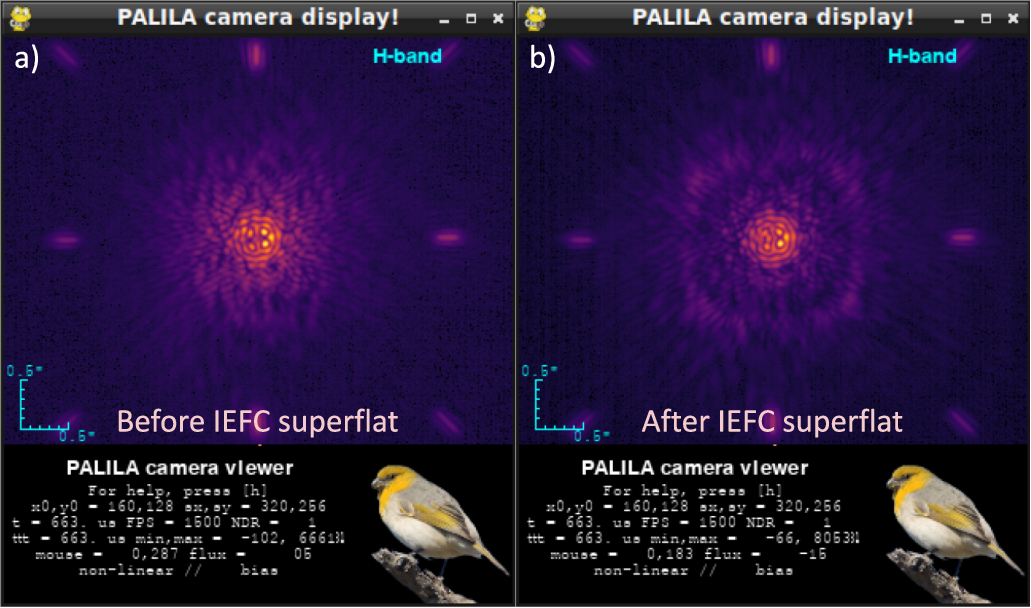}
\end{tabular}
\end{center}
\caption{IEFC used to flatten the deformable mirror: (a) before the superflat map generation, (b) after the superflat map is applied.}
\label{fig:iefc_superflat} 
\end{figure} 

In the calibration step of IEFC, we define a control region around the PSF where the control will try to reduce the flux as much as possible. That region is usually a circular C-shape with inner and outer working angles, or a rectangular shape with a circular notch for the inner working angle. On SCExAO, our pupil is covered by 45 actuators across the diameter, which means that the control region has an outer separation of 22.5~$\lambda/D$. Due to a coding mistake, we ended up defining a fully circular dark hole and realized that the IEFC control loop was converging at first, before diverging. This is because we have amplitude errors left over from a slightly incorrect flat map on the DM.

We used IEFC to correct those amplitude errors, by stopping the control before the correction goes too far. The result is presented in Fig.~\ref{fig:iefc_superflat}. The contrast was improved around the coronagraphic mask, but a ring of speckles appeared close to the edges of the control region. This is an acceptable result, as good contrast at small separations is more important than at about 1~arcsec. In the future, we might try to improve this result by iterating between two different control regions, one focusing on smaller separations, and one improving the larger separations.

\subsection{IEFC For On-Sky Speckle Control}
\label{sec:IEFC_onsky}

\begin{figure} [b]
\begin{center}
\begin{tabular}{c}
\includegraphics[width=0.97\textwidth]{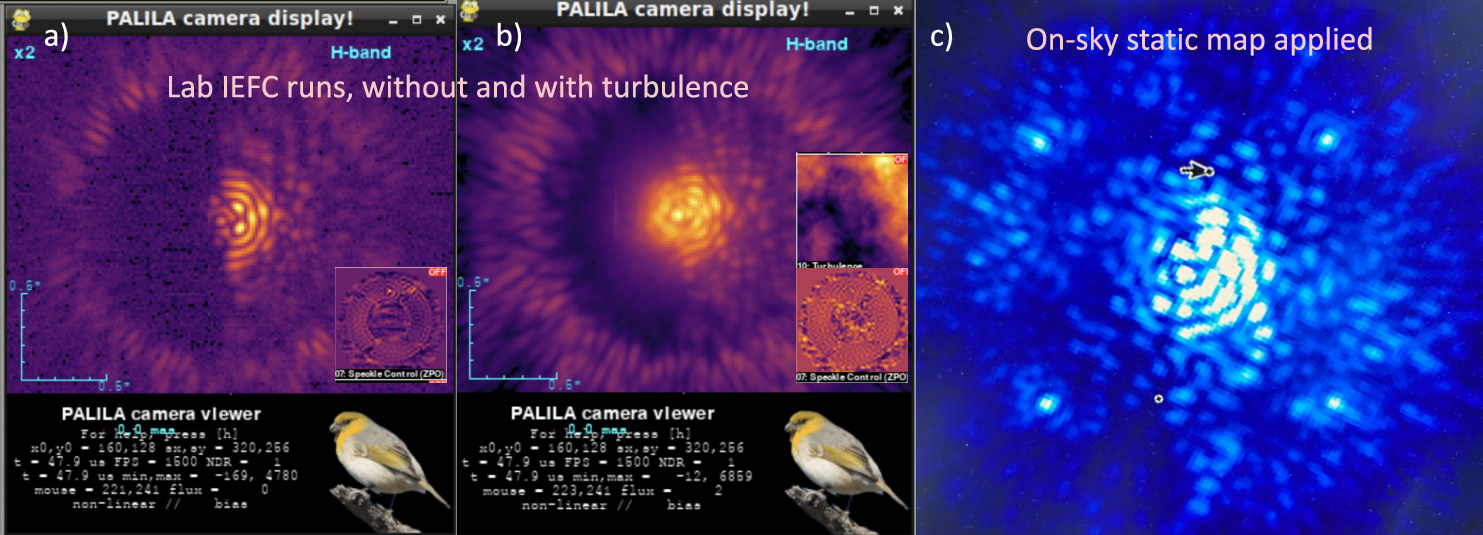}
\end{tabular}
\end{center}
\caption{Towards an on-sky use of IEFC: (a) IEFC map generation on the internal source, without disturbances. (b) IEFC map generation while applying simulated turbulence on the DM, and correcting it with the PyWFS. (c) example of CHARIS frame taken with a static dark hole map generated in the lab, and applied on-sky. The contrast improvement is very marginal (factor $\sim$2).}
\label{fig:iefc_results} 
\end{figure} 

SPIDERS demonstrated that it is possible to dig dark holes on top of AO3k residuals. We used IEFC to dig dark holes in several conditions: In narrowband mode around 1.55~\mum, or on the whole H-band, using both a C-RED 2 (Palila) and a C-RED ONE camera (Apapane), by controlling the DM directly and by applying zero-point offsets on the DM, in the presence of simulated turbulence (see Fig.~\ref{fig:iefc_results} (a) and (b)). In lab settings, even when turbulence is added, IEFC converges to a good dark zone, although the contrast at small inner working angles is affected. But once we moved to on-sky testing, things did not go well. As we demonstrated in the past, applying a static map calculated beforehand improves the contrast marginally, by maybe a factor 2 (see Fig.~\ref{fig:iefc_results} (c)). But running the algorithm on-sky has so far been unsuccessful. The limited amount of time on-sky was disturbed with telescope vibrations and LWE, which created unstable conditions for proper testing. So far, we tried using a lab calibration to close the loop on-sky, unsuccessfully. We also tried to calibrate on-sky, but even on a bright target, with more than a thousand modes to probe, it takes 20 to 45~minutes to calibrate, during which the conditions should not vary. Again this did not work so far. Using a lab calibration would be ideal in practice, so we will focus on understanding how to improve on it for on-sky testing.

\section{CONCLUSION AND PERSPECTIVE}

With the AO3k and NBS upgrades, the AO instrumentation on the NasIR platform of Subaru has significantly improved in the last couple of years, in wavefront correction and operation efficiency. We demonstrated extreme-AO performance with AO3k, thanks to the NIRWFS, which received some useful upgrades last summer. Notably, the new optical binning mode, or low-resolution mode, allows us to reach even fainter NIR targets that were only accessible with a laser guide star in the past.

\begin{figure} [b]
\begin{center}
\begin{tabular}{c}
\includegraphics[width=0.6\textwidth]{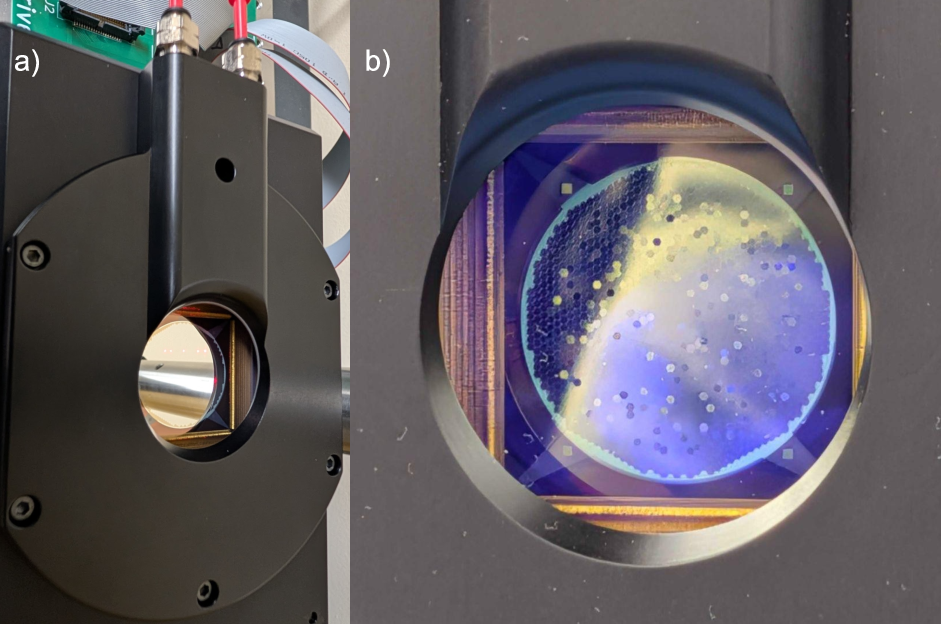}
\end{tabular}
\end{center}
\caption{New deformable mirror that will replace the current 2k MEMS DM. The new DM is a segmented 3k MEMS DM (1022 segments). (a) Device shipped to SCExAO. (b) First device built by BMC and stress tested: Some segments got stuck at weird angles. This is now a display item shown during SPIE.}
\label{fig:bmc_hex3k} 
\end{figure} 

Thanks to the first stage Ao being an extreme-AO now, we hope to use SCExAO to focus on speckle control at small inner working angles. This was demonstrated by the visiting technology demonstrator SPIDERS, installed in the visitor port of the NBS. Unfortunately, we still suffer from two major issues: telescope vibrations and low-wind effect. Telescope vibrations, only visible by SCExAO, forces the SCExAO control loop to correct large tip/tilt errors instead of focusing on precise speckle correction. Similarly, Subaru's thick spiders induce LWE modes that are difficult to correct with the limited stroke of the MEMS DM. New control loops are being implemented, notably Fast\&Furious for non-coronagraphic images, and the LLOWFS when the coronagraph is used, show promising results to tackle these issues.

Although we have been testing speckle control for a long time at this point, we are finally getting to a point where these algorithms can be tested on-sky. We are now focusing on digging dark holes with IEFC, which works well in various conditions in lab settings. On-sky testing has not been successful so far, but we remain hopeful we can achieve the level of stability required to succeed.

In addition to our work in improving speckle control and stability, we are also improving SCExAO's hardware. In the next year, we are planning to implement more efficient coronagraphs, as well as replace our current 2k MEMS DM with a 3k segmented MEMS DM (see Fig.~\ref{fig:bmc_hex3k}). This will allow us to have more actuators in the pupil (55 across the pupil instead of 45), but also test new wavefront control for segmented apertures, like the future Thirty Meter Telescope (TMT), or space missions like the Habitable Worlds Observatory.

\acknowledgments 
Based on data collected at Subaru Telescope, which is operated by the National Astronomical Observatory of Japan. The development of SCExAO and AO3k is supported by the Japan Society for the Promotion of Science (Grant-in-Aid for Research \#23340051, \#26220704, \#23103002, \#19H00703, \#19H00695, and \#21H04998), Subaru Telescope, the National Astronomical Observatory of Japan, the Astrobiology Center of the National Institutes of Natural Sciences, Japan, the Mt Cuba Foundation, and the Heising-Simons Foundation. CHARIS was built at Princeton University under a Grant-in-Aid for Scientific Research on Innovative Areas from MEXT of the Japanese government (\# 23103002). The development of the CACAO software is supported by the National Science Foundation under award 2410616. Maunakea is a mountain that connects Kanaka Maoli (Native Hawaiians) to their ancestral and universal origins. We, as scientists and educators, acknowledge this interconnectedness between Kanaka and their `\=Aina (land).  We recognize that we have a kuleana (responsibility) that comes with the opportunity to pursue astronomical research on Maunakea. It is our kuleana to build and strengthen healthy relationships with the land and the people of this place based on mutuality and trust.

\bibliography{report} 
\bibliographystyle{spiebib} 

\end{document}